\documentclass[a4paper,12pt,reqno]{amsart}
\usepackage{amsmath}%
\usepackage{amsfonts}%
\usepackage{amssymb}%
\usepackage{graphicx,epstopdf,caption}
\usepackage{enumerate}
\usepackage{amscd,amssymb,epsf}
\usepackage{amsfonts,latexsym}
\usepackage{epsf,graphicx}
\usepackage{subfigure,wrapfig}
\usepackage{algorithmic}
\newtheorem{theorem}{Theorem}
\theoremstyle{plain}

\numberwithin{equation}{section}

\newcommand{\R}{{\mathbb R}}
\newcommand{\C}{{\mathbb C}}

\newcommand{\sech}{\mathop{\rm sech}\nolimits}
\newcommand{\dbar}{\overline{\partial}}

\newcommand{\divergence}{\mathop{\rm div}\nolimits}

\newcommand{\lp}{\left(}
\newcommand{\rp}{\right)}
\renewcommand{\Vec}[1] {\left( \begin{array}{c} #1 \end{array}\right)}
\newcommand{\Cases}[1] {\left\{ \begin{array}{ll} #1 \end{array}\right.}
\newcommand{\Matrix}[2]{\left[\begin{array}{*{#1}{c}} #2 \end{array}\right]}
\newcommand{\pr}{{\prime}}

\begin{document}
\title[Transverse instability of  solutions of the NV equation]
{Transverse instability of plane wave soliton solutions of the
  Novikov-Veselov equation}

\author{R. Croke, J. L. Mueller and A. Stahel}

\begin{abstract}
The Novikov-Veselov (NV) equation is a dispersive (2+1)-dimensional
nonlinear evolution equation that generalizes the (1+1)-dimensional
Korteweg-deVries  (KdV) equation.  This paper considers the  stability
of plane wave soliton solutions of the NV equation to transverse
perturbations. To investigate the behavior of the perturbations, a
hybrid semi-implicit/spectral numerical scheme was developed,
applicable to other nonlinear PDE systems.  Numerical simulations of
the evolution of transversely perturbed plane wave solutions are presented.
In particular, it is established that plane wave soliton solutions are
not stable for transverse perturbations.
\end{abstract}
\maketitle

\section{Introduction}
The Novikov-Veselov (NV) equation for $u(z,t) = u(x,y,t)$ was
introduced in the periodic  setting by Novikov and Veselov
\cite{veselovnovikov} in the form
$$u_t = -\partial_z^3 u -\dbar^3u +3\partial_z(uf)
+3\dbar_z(u\bar{f}), \quad \mbox{where} ~~ \dbar_z f=\partial_z u,$$ 
with $\dbar_z=\frac{1}{2}(\partial_x +i\partial_y)$,   
$\partial_z=\frac{1}{2}(\partial_x -i\partial_y)$, 
where it was derived algebraically from a Lax triple, and from this
point of view is considered the most general derivation of the KdV
equation \cite{Bogdanov1987}.

While there are no known physical applications of the NV equation, it
is related to two other (2+1)-dimensional integrable systems which
have been more widely studied. The Davey-Stewartson II (DS II)
equation describes the complex amplitude of surface waves in shallow
water 
\[i u_t = -2\partial_z^2 u -2\dbar_z^2u 
- \dbar_z\partial_z^{-1}|u|^2  - \partial_z^{-1}\dbar_z|u|^2,\]
and was proved in \cite{Sung1994} to be completely integrable.
The modified NV equation (mNV) is a member of the DS II hierarchy:
\begin{eqnarray*}
u_t &=& -\partial_z^3 u -\dbar^3u 
   -\frac{3}{4}(\partial_z\bar{u})(\dbar_z\partial_z^{-1}|u|^2 ) 
   - \frac{3}{4}(\dbar_z u)(\dbar_z\partial_z^{-1}|u|^2 ) \\
&&- \frac{3}{4}\bar{u}\dbar_z\partial_z^{-1}(\bar{u}\dbar_z u) 
   -\frac{3}{4}u\partial_z^{-1}(\dbar_z(\bar{u}\dbar_z u).
\end{eqnarray*}
Here $\dbar_z^{-1}$ and $\partial_z^{-1}$ are the solid Cauchy
transforms defined by
\begin{equation*}
\dbar_z^{-1}u = \frac{1}{\pi}\int_{\R^2}\frac{1}{z-\zeta}d\zeta, \quad 
\partial_z^{-1}u = \frac{1}{\pi}\int_{\R^2}
    \frac{1}{\bar{z}-\overline{\zeta}}d\zeta
\end{equation*}

The integrability of the NV equation has been recently proved in
\cite{Perry2012} where it is shown that a Miura-type map takes
solutions of the mNV equation to solutions of the NV equation with
initial data of conductivity type.  This type of initial condition for
the NV equation was first studied in \cite{LMS2005} where it was shown
that the inverse scattering method for the NV equation is well-posed
for initial conditions of conductivity type.  In \cite{LMSS_2012a} it
was shown that an initially radially-symmetric conductivity-type
potential evolved under the ISM does not have exceptional points and
is itself of conductivity-type.  In \cite{LMSS_2012b}  evolutions of
rotationally symmetric, compactly supported initial data of
conductivity type computed from a numerical implementation of the
inverse scattering method for NV are compared to evolutions of the NV
computed from a semi-implicit finite-difference discretization of NV
and are found to agree with high precision.  This supported the
integrability conjecture that was then established in \cite{Perry2012}
where the class of initial data was enlarged by applying Miura-map
techniques rather than the scattering maps studied in
\cite{LMS2005,LMSS_2012a,LMSS_2012b}.  In \cite{MusicPerrySiltanen2013} it is shown that the set of conductivity type potentials is unstable under $C^{\infty}_0$ perturbations.

 If we consider real solutions $u(x,y,t)$ and let $f(x,y,t) = v(x,y,t)
 + i\,w(x,y,t)$, the NV equation has an equivalent representation in
 $(x,y)$-space:
 \begin{eqnarray}
 4\, u_t &=& - u_{xxx} + 3\,u_{xyy} + 3\,(uv)_x + 3\,(uw)_y, \label{eq:NVreal}\\
 u_x &=& v_x - w_y, \label{Aux1}  \\ 
 u_y &=& -w_x - v_y.\label{Aux2}
 \end{eqnarray}
If the functions $u,v$ and $w$ are not dependent on $y$, the NV
equation reduces to a KdV-type equation 
\[ 0=4\,u_t +u_{xxx} - 6\,u\,u_x,\qquad u=v,\qquad w=0\]
and admits soliton solutions of the form
\begin{eqnarray}
 u(x,y,t) &=& -2\,\mbox{c}\,\sech^2(\sqrt{c}(x - c\,t)),\label{eq:Sech2Sol1}\\
 v(x,y,t) &=& -2\,\mbox{c}\,\sech^2(\sqrt{c}(x - c\,t)),\label{eq:Sech2Sol2}\\
 w(x,y,t) &=& 0. \label{eq:Sech2Sol3} 
\end{eqnarray}

To facilitate the investigation of the qualitative nature of solutions
to the NV equation,  we present a version of a semi-implicit
pseudo--spectral numerical scheme introduced by Feng et al
\cite{Feng1999} that solves the Cauchy problem for the NV equation.
This constitutes the first numerical implementation of a spectral
method for a system of soliton nonlinear PDE's. It is shown to
preserve the $L_2$ norm for the KdV type equation and is considerably
faster and requires less computer memory allocation than the finite
difference scheme introduced in~\cite{LMSS_2012b} for the solution of
the Cauchy problem. In~\cite{Allen2007} the method in~\cite{Feng1999} was
applied to the KP equation to investigate the stability of soliton
solutions.  Here, we adapt the method for systems of equations to
study numerically the nature of the instability of traveling wave
solutions to the NV equation to transverse perturbations.

The problem of transverse stability of traveling wave solutions has
been studied for many of the classic soliton equations including the
KP equation \cite{AllenRowlands1997,Allen2007,Bridges2001,Infeld1999},
the Boussinesq equation~\cite{AllenRowlands1997}, the ZK equation
\cite{AllenRowlands1993,FryczInfeld1989,Infeld1985,InfeldFrycz1987},
and most notably, the KdV equation \cite{KP1970}.  For the NV
equation, we carry out a linear stability analysis by considering
sinusoidal perturbations with wavefront perpendicular to the direction
of propagation.  In order to draw conclusions about the instability of
soliton solutions, as well as approximate the growth rate, we apply the method
developed by Rowlands, Infeld, and Allen
\cite{AllenRowlands1993,InfeldRowlands2000}.
Due to the complicated boundary conditions, we employ a geometric
optics limit based on a scheme that assumes the nonlinear wave
undergoes a long-wavelength perturbation.  Thus, if the wave vector of
the perturbation is ${\bf k}$, we assume it is very small in
comparison to the wave vector of the solution.  This type of
investigation is known as the ${\bf K}$-expansion method.    To our
knowledge, the only other use of this method for a soliton system is
in the work of Bradley \cite{Bradley1999} in a model of small
amplitude long waves traveling over the surface of thin
current-carrying metal film.

In \cite{AllenRowlands1993}, the authors conjecture that to do a
linear stability analysis using the ${\bf K}$-expansion method, a
regular perturbation analysis is consistent if and only if the
equation is an integrable system.  If a multiscale analysis is needed,
the equation is not integrable.  The ZK equation and the equation in
\cite{Bradley1999} are not integrable, and the ordinary perturbation
analysis fails.  So far this conjecture is supported by the KP and
Boussinesq equations \cite{AllenRowlands1997}.  The results here
support the conjecture by showing that only an ordinary perturbation
analysis is needed for the NV equation. 


The paper is organized as follows.  In Section 2 we show that planar
solutions to the NV equation must be a solution of a KdV-type
equation.  In Section 3 the semi-implicit pseudo-spectral method is
presented, with its linear stability analysis in Section 3.1.  The
${\bf K}$-expansion method is used to establish the instability of
traveling wave solutions of NV to transverse perturbations in Section~4.
Numerical results are found in Section~5 and conclusions in Section~6.

\section{From planar solutions of Novikov--Veselov to KdV}
We examine planar solutions to the Novikov--Veselov
equations~\eqref{eq:NVreal}--\eqref{Aux2}, i.e. solutions that
depend only on one spatial variable $s=(n_1,n_2)\cdot(x,y)$,
moving in a direction given by the vector 
$\vec n=(n_1,n_2)=(\cos(\alpha),\sin(\alpha))$. We seek solutions of
the form
\begin{eqnarray*}
  u(t,s) &=& u(t,x,y) = u(t,n_1\,s, n_2\, s)\\
  v_i(t,s) &=& v_i(t,x,y) = v_i(t,n_1\,s, n_2\, s)\quad\mbox{for }i=1,2\\
 u^\pr(t,s) &=& \frac{\partial}{\partial s}\, u(t,s) 
      = n_1 u_x(t,x,y) + n_2 u_y(t,x,y)
\end{eqnarray*}
The assumption that $u$, $v_1$ and $v_2$ are independent on
$n_2\,x-n_1\,y$ is equivalent to
\[   n_2 u_x -n_1u_y =
   n_2\,\frac{\partial\,v_1}{\partial x} -n_1\,\frac{\partial\,v_1}{\partial y} =
   n_2\,\frac{\partial\,v_2}{\partial x} -n_1\,\frac{\partial\,v_2}{\partial y} =0 \]
 As a consequence we obtain
$ u_x = n_1\,u^\pr$ and $q_x = n_2\,u^\pr$. 
The goal is to find a PDE for $u(t,s)$.  For a given $u(t,s)$ the $\bar\partial$ equation
\[\left\{ \begin{array}{rcl} 
  \frac{\partial}{\partial x}\,v_1 -\frac{\partial }{\partial y}\,v_2  &=& +u_x \\
  \frac{\partial}{\partial x}\,v_2 +\frac{\partial }{\partial y}\,v_1  &=& -u_y
  \end{array}\right.\]
translates to
\[\left\{ \begin{array}{rcl} 
  n_1\,v_1^\pr -n_2\,v_2^\pr  &=& +n_1\,u^\pr \\
  n_2\,v_1^\pr +n_1\,v_2^\pr  &=& -n_2\,u^\pr
  \end{array}\right.\quad\Longrightarrow\qquad
\begin{array}{rcrcl}
 v_1^\pr &=& (n_1^2 -n_2^2) &u^\pr\\
 v_2^\pr &=& -(2\,n_1\,n_2^2)& u^\pr  
\end{array}\]
with the solutions
\begin{eqnarray*}
 v_1(t,s) &=& (n_1^2 -n_2^2)\; u(t,s)+c_1\\
 v_2(t,s) &=& -(2\,n_1\,n_2^2)\;u(t,s)+c_2  
\end{eqnarray*}
The nonlinear expression in~\eqref{eq:NVreal} leads to
\begin{eqnarray*}
  \divergence(u\,\vec v)
 &=& (n_1^2-n_2^2)\,(u^2)_x-2\,n_1\,n_2\,(u^2)_y+ c_1u_x + c_2u_y\\
 &=&\left(2\,(n_1^2-n_2^2)\,n_1-4\,n_1\,n_2^2\right)\,u\,u^\pr
 + (c_1\,n_1 + c_2\,n_2)\, u^\pr\\
 &=& 2\,\kappa\,u\,u^\pr + \beta\, u^\pr 
\end{eqnarray*}
where
\begin{eqnarray*}
  \beta =c_1\,n_1 + c_2\,n_2 \hspace{1em} \mbox{and} \hspace{1em} \kappa=\kappa(\alpha) = \,(n_1^2-n_2^2)\,n_1-2\,n_1\,n_2^2 = \cos(3\,\alpha)
\end{eqnarray*}
The factor $\kappa=\cos(3\,\alpha)$ is further evidence for the
threefold rotational invariance of the Novikov--Veselov equations.
Now we examine the Novikov--Veselov equation
\begin{eqnarray*}
u_t &=& -\frac{1}{4}u_{xxx} +\frac{3}{4}u_{xyy}+ \frac{3}{4}\,\divergence(u\,\vec v)\\
 &=&  -\frac{1}{4}\, n_1^3\,u^{\pr\pr\pr}
    +\frac{3}{4}\,n_1\,n_2^2\, u^{\pr\pr\pr}
    +\frac{6}{4}\,\kappa\,u\,u^\pr + \frac{3}{4}\,\beta\,u^\pr\\
 &=&  -\frac{1}{4}\,\kappa\, u^{\pr\pr\pr}
    +\frac{6}{4}\,\kappa\,u\,u^\pr + \frac{3}{4}\,\beta\,u^\pr
\end{eqnarray*}
Thus, a planar solution of the  Novikov--Veselov has to be a solution
of the KdV-like equation
\begin{equation}
  \label{eq:NVKdV}
  \frac{4}{\kappa}\, u_t =  - u^{\pr\pr\pr} + 6\,u\,u^\pr
  +\frac{3\,\beta}{\kappa}\,u^\pr 
\end{equation}
The only essential modification is the contribution proportional to
$u^\pr$ and thus it should not come as a surprise that the solutions
are related.
If $q(t,s)$ is a solution of the standard KdV equation
\begin{equation}
  \label{eq:KdV}
\dot q(t,x) = - q^{\pr\pr\pr}(t,x) + 6\, q(t,x)\,q^\pr(t,x)
\end{equation}
then
\[  u(t,s) = q(\frac{\kappa}{4}\, t, s+\frac{3\,\beta}{4}\,t)\]
is a solution of the Novikov--Veselov equation~\eqref{eq:NVreal}, which can be verified as follows:
\begin{eqnarray*}
u_t &=& \frac{\kappa}{4} q_t + \frac{3\,\beta}{4} q^\pr\\
\frac{4}{\kappa} u_t + u^{\pr\pr\pr}-6\,u\,u^\pr-\frac{3\,\beta}{\kappa}\,u^\pr
&=& q_t + \frac{3\,\beta}{\kappa}\,\,q^\pr + q^{\pr\pr\pr} 
   -6\,q\,q^\pr -\frac{3\,\beta}{\kappa}\,q^\pr\\
&=& q_t + q^{\pr\pr\pr} - 6\,q\,q^\pr\,=\,0
\end{eqnarray*}
For a solution $q(t,s)$ of KdV choose constants $k_1$ and $k_2$ with
\[  \frac{3\,\beta}{4}=k_1 + k_2\,\frac{3\,\kappa}{2}\]
and verify that
\[ u(t,s) = q(\frac{\kappa}{4}\, t\,,\,s+k_1\,t) - k_2\]
is a solution of the Novikov--Veselov equation:
\begin{eqnarray*}
u_t &=& \frac{\kappa}{4} q_t + k_1\,q^\pr\\
\frac{4}{\kappa}\,u_t + u^{\pr\pr\pr}-6\,u\,u^\pr-\frac{3\,\beta}{\kappa}\,u^\pr
&=& q_t+ \frac{k_1\,4}{\kappa}\,q^\pr + q^{\pr\pr\pr} 
   -6\,(q-k_2)\,q^\pr -\frac{3\,\beta}{\kappa}\,q^\pr\\
&=& q_t + q^{\pr\pr\pr} - 6\,q\,q^\pr 
  +\left(\frac{k_1\,4}{\kappa}+6\,k_2-\frac{3\,\beta}{\kappa}\right)\,q^\pr\\
&=& q_t + q^{\pr\pr\pr} - 6\,q\,q^\pr 
  +\frac{4}{\kappa}\,\left(k_1+ k_2\,\frac{3\,\kappa}{2}-\frac{3\,\beta}{4}\right)\,q^\pr\,=\,0
\end{eqnarray*}
{\bf Remarks:}
\begin{itemize}
\item Consider the choice $\beta =k_1=k_2=0$. Then $q(t,s) =
  u(\frac{\kappa}{4}\, t, s)$ 
  is the KdV solution where the time scale is multiplied with
  $\kappa/4$. Since $\kappa=\kappa(\alpha)=\cos(3\,\alpha)$ the speed
  of the solution changes according to an angle dependent profile, as
  shown in Figure~\ref{fig:NVSpeedProfile}.
\begin{figure}[ht]
  \centering
    \includegraphics[width=60mm]{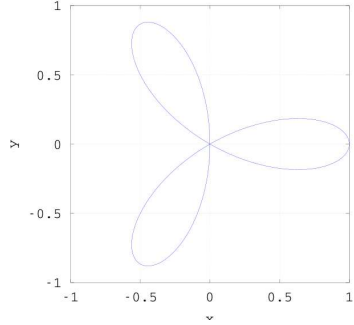}
    \caption{Speed profile for planar solutions of the
      Novikov--Veselov equation}
    \label{fig:NVSpeedProfile}
\end{figure}
\item The additive constant $k_2$ moves the KdV solution up or down in
  the graph.
\item Replacing $s$ by $s+k_1\,t$ (with $k_1 =
  \frac{3\,\beta}{4}-k_2\frac{3\,\kappa}{2}$) corresponds to observing 
  the KdV solution in a moving frame, where the frame moves with
  velocity $-k_1$.
\end{itemize}

\section{A Pseudo-Spectral method for the solution of (2+1) 
nonlinear  wave equations}
\label{sec:Spectral}

To numerically solve~\eqref{eq:NVreal}--\eqref{Aux2} we use a
semi--implicit leap--frog spectral method based on the method in
\cite{Feng1999}.  We restrict ourselves to a finite spatial domain
$\Omega = [0,W_x]\times[0,W_y]$ with periodic boundary conditions.
Thus we work on a torus topology and for the numerical results we have
to observe the traveling waves across the boundary. We must also choose
the domain large enough for the problem at hand.

This method uses the Fast Fourier Transform (FFT) to compute the
spatial evolution and a leapfrog scheme to compute the time-stepping.
The numerical scheme can be summarized as follows:

The Fourier transform of \eqref{eq:NVreal}--\eqref{Aux2} is
\begin{eqnarray}
 4\,\hat{u}_t&=& i\,\lp \xi^3-3\,\xi\,\eta^2\rp\,\hat{u}
+3\,i\,\xi\,\mathcal{F}[(uv)] + 3\,i\,\eta\,\mathcal{F}[(uw)],\label{eq:FTNV}\\ 
\xi\,\hat{u} &=& \xi\,\hat{v} - \eta\,\hat{w}, \label{eq:NVAuxFFT1}\\
\eta\,\hat{u}&=& -\eta\,\hat{v} - \nu\,\hat{w}. \label{eq:NVAuxFFT2}
\end{eqnarray}
As usual, $\hat{u}$ refers to the Fourier transform $F$ of $u$ and is given by 
\begin{eqnarray*}
\hat{u}_{p,q} & =& \mathcal{F}[u_{l,m}] 
   = \sum_{l=0}^{L-1}\sum_{m=0}^{M - 1}u_{l,m} \,e^{-i(\xi_px_l + \eta_qy_m)},\\
u_{l,m} &=& \mathcal{F}^{-1}[\hat{u}_{p,q}] 
  = \frac{1}{L\,M}\, \sum_{p=-L/2}^{L/2 - 1}\sum_{q=-M/2}^{M/2 -1}\hat{u}_{p,q}\,
      e^{i(\xi_px_l + \eta_qy_m)}. 
\end{eqnarray*}
The parameters $L$ and $M$ are the number of grid points in
$[0,W_x]\times[0,W_y]$, respectively.  They need to be powers of two
in order to use the standard FFT.  The spatial grid is defined by
$(x_l,y_m) = (l\Delta x,m\Delta y)$ where $l = 0,\ldots,L - 1 $, and
$m = 0,\ldots,M - 1$.  The spectral variables are $\lp\xi_p,\eta_q\rp
= \lp 2\pi p/W_x, 2\pi q/W_y\rp$, with $p = -L/2,\ldots$,
$-1,0,1,\ldots,L/2$, and $q =  -M/2,\ldots$, $-1,0,1,\ldots,M/2$.

Equations \eqref{eq:NVAuxFFT1} and \eqref{eq:NVAuxFFT2} can be solved
in terms of $\hat{u}$
\begin{equation}
  \label{eq:NVAuxFFT}
   \hat{v}  = \frac{\xi^2 - \eta^2}{\xi^2 + \eta^2}\,\hat{u},\qquad
   \hat{w}  = \frac{-2\eta\,\xi}{\eta^2 + \xi^2}\,\hat{u}
\end{equation}
Special attention has to be paid to the case $\xi^2+\eta^2=0$. The
corresponding Fourier coefficient represent the average values of the
functions $v$ and $w$.

Let ${\vec c} (t)$ be the vector with the $LM$ Fourier coefficients of
the solution $u(x,y,t)$. Then solving~\eqref{eq:NVAuxFFT1},
\eqref{eq:NVAuxFFT2} and computing $ -3\,i\,(\xi\,\mathcal{F}[(uv)] 
+ \eta\,\mathcal{F}[(uw)])$ may be written as one nonlinear
function $\vec F({\vec c} )$. With the well constructed diagonal matrix
$\mathbf{D}$ the NV equation~\eqref{eq:FTNV} reads as
\begin{eqnarray}
  \label{eq:NVFTODE}
  4\,\frac{d}{dt}\, {\vec c} (t) = \mathbf{D}\, {\vec c} (t) + \vec F({\vec c} (t))
\end{eqnarray}

%
For the time integration we use a symmetric three-level difference
method for the linear terms, and a leapfrog  method for the nonlinear
terms. For the parameter $0\leq \theta\leq 1$ we use a superscript to
refer to the time iteration and the above leads to
\begin{eqnarray}
  \frac{4}{2\,\Delta t} \,({\vec c}\,^{n+1}-{\vec c}\,^{n-1})
&=& \theta\,\mathbf{D}\,(\vec c\,^{n+1}+\vec c\,^{n-1}) +
(1-2\,\theta)\,\mathbf{D}\, \vec c\,^n + \vec F(\vec c\,^n)\notag\\
(2-\theta\,\Delta t\,\mathbf{D})\,\vec c\,^{n+1} &=&
(1-2\,\theta)\,\Delta t\,\mathbf{D}\, \vec c\,^n
+(2+\theta\,\Delta t\,\mathbf{D})\,\vec c\,^{n-1}\label{eq:NVFTODEnum}\\
&&+\Delta t\, \vec F(\vec c\,^n)\notag
\end{eqnarray}
Thus we have an implicit scheme for the linear contribution and an
explicit scheme for the nonlinear contribution.
Since the matrix $\mathbf{D}$ is diagonal we do not have to solve a
system of linear equations at each time step.
For the special case $\theta = 1/2$ we obtain a Crank Nicolson scheme
\[(4-\Delta t\,\mathbf{D})\,\vec c\,^{n+1} =
+(4+\Delta t\,\mathbf{D})\,\vec c\,^{n-1}+2\,\Delta t\, \vec F(\vec c\,^n)\]



For  the three level method we need a separate method for the first
time step. For the sake of simplicity we may choose $\vec c\,^{-1}=\vec c\,^0$.
For computations with known solutions, e.g.~\eqref{eq:Sech2Sol1}, we
use the known values of the solution at time $-\Delta t$.



%

\subsection{Linear Numerical Stability Analysis}

To gain insight into the stability of the spectral method, we include
a linear stability analysis.  The parameters determined here were used
in the numerical experiments that follow.   We examine the problem on a domain
$\Omega = [0,W]\times[0,W]\subset \R^2$ and use periodic boundary conditions.
Thus we examine the initial boundary value problem
\begin{equation}
    \label{eq:NVlinearized}
\begin{array}{l} 
  4\,u_t = -u_{xxx} + 3\,u_{xyy} + 3\,\alpha\,(v_x + w_y),\\
 u_x = v_x - w_y,\qquad u_y = -w_x - v_y
\end{array}
\end{equation}
The case $\alpha=0$ is the linearization of~\eqref{eq:NVreal}--\eqref{Aux2}
about the zero solution.
The utility of a linear stability analysis for a nonlinear
system was addressed in \cite{Feng1999} in the context of the KP and
ZK equations, where they found that their results for the linear
stability analysis were validated by numerical results and argued that
while the analysis does not prove stability and convergence of the
nonlinear scheme, the obtained stability conditions often suffice in
practice.
When implemented on a domain with periodic boundary
condition~\eqref{eq:NVlinearized} leads to
  \begin{equation}
    \label{eq:NVlinearizedPeriodic}
\begin{array}{ll} 
  4\,u_t = -u_{xxx} + 3\,u_{xyy} + 3\,\alpha\,(v_x + w_y),
  & (x,y)\times t\in \Omega\times \R,\\
 u_x = v_x - w_y,\qquad u_y = -w_x - v_y&\\
u(x,y,0) = u_0(x,y), & (x,y) \in \Omega,\\
u(x,y,t) = u(x + W_x,y,t), & (x,y)\times t\in \R^3,\\
u(x,y,t) = u(x,y + W_y,t), & (x,y)\times t\in \R^3,
\end{array}
  \end{equation}
Using Fourier series and~\eqref{eq:NVAuxFFT} this leads to
\begin{eqnarray*}
  4\,\hat u_t &=& i\,(\xi^3-3\,\xi\,\eta^2)\,\hat u 
+ 3\,i\,\xi\,\alpha\,\frac{\xi^2-\eta^2}{\eta^2 + \xi^2}\,\hat{u}
+ 3\,i\,\eta\,\alpha\,\frac{-2\eta\,\xi}{\eta^2 + \xi^2}\,\hat{u}\\
&=& i\,\omega\,\hat u + i\,\frac{3\,\omega\,\alpha}{\eta^2 + \xi^2}\,\hat{u}
\end{eqnarray*}
where $\omega = \xi^3-\xi\,\eta^2$. This is an ODE of the form
\begin{equation}
  \label{eq:ODE}
   4\,\frac{d}{dt}\, u(t) = i\,\lambda\, u(t) + i\,\gamma\, u(t)
\end{equation}
where $\lambda=\omega=\xi^3-3\,\xi\,\eta^2$ and
$\gamma=\omega\,\frac{3\,\alpha}{\eta^2+\xi^2}$.
Using the numerical scheme~\eqref{eq:NVFTODEnum} this leads to
\[  (2-i\,\theta\,\lambda\,\Delta t)\, u^{n+1}
  = i\,(1-2\,\theta)\,\lambda\,\Delta t\, u^{n}
  + (2+i\,\theta\,\lambda\,\Delta t)\, u^{n-1}
  +i\,\gamma\,\Delta t\, u^n\]
or with
\begin{equation}
  \label{eq:StabCoeff}
   b_1 = 2+i\,\theta\,\lambda\,\Delta t\,\in\C,\qquad
   b_2 = i\,((1-2\,\theta)\,\lambda+\gamma)\,\Delta t\in\C
\end{equation}
to the iteration matrix
\[\Vec{u^{n+1}\\u^n} = \mathbf{M}\,\Vec{u^{n}\\u^{n-1}}
 =\Matrix{2}{\frac{b_2}{\overline{b_1}}&\frac{b_1}{\overline{b_1}}\\1&0}
  \, \Vec{u^{n}\\u^{n-1}}\]
For the system to be stable we have to verify that the norm of the
eigenvalues $z_{1,2}$ of $\mathbf{M}$ are less or equal to~1. The
characteristic equation is given by
\[\Phi_{NV}(z)= z^2 -\frac{b_2}{\overline{b_1}}\,z -\frac{b_1}{\overline{b_1}}=0\]
Thus we conclude $|z_i\cdot z_2| = |\frac{b_1}{\overline{b_1}}|=1$ and
stability of the solution is equivalent to $|z_{1,2}|=1$~, i.e. we
have $|u^{n+1}| = |u^{n-1}|$.
Using the explicit solution for quadratic equations and
\begin{equation}
  \label{eq:StabCond}
   b_2\in i\,\R ,\qquad |b_2|\leq 2\,|b_1|
\end{equation}
we find
\begin{eqnarray*}
  2\,\overline{b_1}\, z_{1,2} &=&  b_2\pm\sqrt{b_2^2+4\,\overline{b_1}\,b_1}
 \,=\,  b_2\pm\sqrt{b_2^2+4\,|b_1|^2}\\
 |2\,\overline{b_1}|^2\, |z_{1,2}|^2
&=& 
  |b_2|^2 -|b_2|^2+4\,|b_1|^2 \,=\, |2\,b_1|^2
\end{eqnarray*}
and consequently $|z_{1,2}|=1$~. With~\eqref{eq:StabCoeff} the
stability condition~\eqref{eq:StabCond} for the ODE~\eqref{eq:ODE} is
  \begin{eqnarray*}
   ((1-2\,\theta)\,\lambda+\gamma)^2\,(\Delta t)^2
&\leq& 4\,|2+i\,\theta\,\lambda\,\Delta t\,|^2 \\
 \left(  ((1-2\,\theta)\,\lambda+\gamma)^2
-(2\,\theta\,\lambda)^2\right)\,(\Delta t)^2 &\leq& 16\\
   ((1-4\,\theta)\,\lambda^2+2\,(1-2\,\theta)\,
  \lambda\,\gamma+\gamma^2)\,(\Delta t)^2 &\leq& 16
  \end{eqnarray*}
For $\gamma=\alpha=0$ and $\frac{1}{4}\leq \theta\leq 1$ the stability
condition is satisfied, independent on the step size $0<\Delta t$
and we have a stability result for the initial boundary value
problem~\eqref{eq:NVlinearizedPeriodic}~.

\begin{theorem} 
Apply the numerical scheme~\eqref{eq:NVFTODEnum} to the linearization
of~\eqref{eq:NVreal}--\eqref{Aux2} about the zero solution, i.e. to the equation
\[ \begin{array}{rcl}
 4\,u_t &=& -u_{xxx} + 3\,u_{xyy} \\
 u_x &=& v_x - w_y\;,\qquad u_y = -w_x - v_y  
\end{array}\]
This scheme is stable for $\frac{1}{4}\leq \theta\leq 1$.
\end{theorem} 

Proving stability for~\eqref{eq:NVlinearizedPeriodic} with $\alpha\neq 0$
requires a few more computations.
Use the expressions for $\lambda=\omega=\xi^3-3\,\xi\,\eta^2$ and
$\gamma=\omega\,\frac{3\,\alpha}{\eta^2+\xi^2}$ in~\eqref{eq:ODE}
and the stability condition reads as
\begin{eqnarray*}
   ((1-4\,\theta)\,\lambda^2+2\,(1-2\,\theta)\,
  \lambda\,\gamma+\gamma^2)\,(\Delta t)^2 &\leq& 16\\
  \omega^2\; ((1-4\,\theta)\,+2\,(1-2\,\theta)\,\frac{3\,\alpha}{\eta^2+\xi^2}
    +\frac{9\,\alpha^2}{(\eta^2+\xi^2)^2})\,(\Delta t)^2 &\leq& 16\\
  (1-4\,\theta)\,+2\,(1-2\,\theta)\,\frac{3\,\alpha}{\eta^2+\xi^2}
    +\left(\frac{3\,\alpha}{\eta^2+\xi^2}\right)^2 
  &\leq& \left(\frac{4}{\omega\,\Delta t}\right)^2
\end{eqnarray*}
For $\theta=\frac{1}{2}$ we find the sufficient conditions.
\begin{eqnarray*}
  -2+\left(\frac{3\,\alpha}{\eta^2+\xi^2}\right)^2 
\,\leq \, \left(\frac{3\,\alpha}{\eta^2+\xi^2}\right)^2 
  &\leq& \left(\frac{4}{(\xi^3-3\,\xi\,\eta^2)\,\Delta t}\right)^2\\
(\Delta t)^2 &\leq& \left(
\frac{4\,(\xi^2+\eta^2)}{3\,\alpha\,(\xi^3-3\,\xi\,\eta^2)}\right)^2
\end{eqnarray*}
Since $ |\xi^3-3\,\xi\,\eta^2| \leq (\xi^2+\eta^2)\cdot\max\{|\xi|,\,|\eta|\}$
we have the sufficient condition
\begin{equation}
  \label{eq:StabCondDt}
  \Delta t\leq \frac{4}{3\,|\alpha|\,\max\{|\xi|,\,|\eta|\}}
\end{equation}
For the DFT with $\Delta x =\Delta y = \frac{W_x}{L}$ we find
$\max\{|\xi|,\,|\eta|\}=\frac{\pi}{\Delta x}$ and thus we have the
stability condition
\begin{equation}
  \label{eq:StabCondDt2}
  \Delta t\leq \frac{4}{3\,|\alpha|\,\pi}\;\Delta x
\end{equation}
As a consequence we have a stability result for the initial boundary
value problem~\eqref{eq:NVlinearizedPeriodic} with $\alpha\neq 0$.

\begin{theorem} 
Apply the numerical scheme~\eqref{eq:NVFTODEnum} to
\eqref{eq:NVlinearizedPeriodic}, i.e.
\[ \begin{array}{rcl}
 4\,u_t &=& -u_{xxx} + 3\,u_{xyy} + 3\,\alpha\,(v_x + w_y)\\
 u_x &=& v_x - w_y\;,\qquad u_y = -w_x - v_y  
\end{array}\]
This scheme is stable for $\theta=\frac{1}{2}$ if the time step
$\Delta t$ satisfies condition~\eqref{eq:StabCondDt2}.
\end{theorem}

\section{Instability of traveling-wave solutions of the 
  NV-equation to transverse perturbations} 

The ~ {\bf K}-expansion method presented by Allen and Rowlands considers
long wavelength perturbations in the transversal direction.  It was
originally used to investigate the stability of solutions to the
Zakharov--Kuznetsov equation, a non-integrable generalization of the
KdV equation \cite{Infeld1985,InfeldFrycz1987,InfeldFrycz1989}.  Since
then the method has been applied to the KP equation and the modified ZK (mZK)
equation \cite{MunroParkes1999}.  This is the second application of
the method to a system of PDE's, the first being by
Bradley~\cite{Bradley1999}. 

\subsection{The modified equations}
To begin, we transform the system to move along with the soliton by
using new independent and dependent variables.
Using 
\[ (t,x,y) \mapsto
\sqrt{c}\,(\alpha\,t,x-c\,t,y)=(\tilde{t},\tilde{x},\tilde{y})\]
and
\[ (u,v_1,v_2) \mapsto c\,(u,v_1,v_2)\]
or
\begin{eqnarray}
 u(t,x,y) &=& c\;\tilde{u}(t\,c^{3/2},(x-c\,t)\,c^{1/2},y\,c^{1/2})\nonumber\\
v_1(t,x,y)&=& c\;\tilde{v}_1(t\,c^{3/2},(x-c\,t)\,c^{1/2},y\,c^{1/2})
      \label{eq:Transformations}\\
v_2(t,x,y)&=& c\;\tilde{v}_2(t\,c^{3/2},(x-c\,t)\,c^{1/2},y\,c^{1/2}), \nonumber
\end{eqnarray}
the NV equations~\eqref{eq:NVreal}--\eqref{Aux2} are transformed to
\begin{eqnarray}
  0 &=& 4u_t - 4u_x + u_{xxx} - 3u_{xyy} - 3(uv)_x -  3(uw)_y, 
\label{eq:TransNV}\\ 
 u_x &=& +v_x -  w_y, \label{eq:TransAux1}\\
 u_y &=& -w_x -  v_y \label{eq:TransAux2}
\end{eqnarray}
For sake of a more readable notation we dropped the tildes on the
new dependent and independent variables. The known solution
\[f_c(x-c\,t) = -2\,c\,\sech^2(\sqrt{c}\,(x-c\,t))\]
of the original system turns into a stationary solution of the modified system.
\[ u_0(x)=v_0(x) = -2\,\sech^2(x), \qquad w_0(x)=0\]

Now we examine perturbed functions of the form
\begin{eqnarray}
u(x,y,t) &=& u_0(x) + \epsilon\, f(x)\,e^{iky + \gamma t}, \label{eq:Main1}\\
v(x,y,t) &=& v_0(x) + \epsilon\, g(x)\,e^{iky + \gamma t}, \label{eq:Main2}\\
w(x,y,t) &=& w_0(x) + \epsilon\, h(x)\,e^{iky + \gamma t}. \label{eq:Main3}
\end{eqnarray}
Thus we examine periodic, transversal perturbations with exponential growth.
For numerical purposes we may also work with a purely real formulation.
\begin{eqnarray}
u(x,y,t) &=& u_0(x) + \epsilon\, f(x)\,\cos(k\,y)\,e^{\gamma t}, \label{eq:Main1R}\\
v(x,y,t) &=& v_0(x) + \epsilon\, g(x)\,\cos(k\,y)\,e^{\gamma t}, \label{eq:Main2R}\\
w(x,y,t) &=& w_0(x) + \epsilon\, h(x)\,\sin(k\,y)\,e^{\gamma t}. \label{eq:Main3R}
\end{eqnarray}

Substituting these expressions into~\eqref{eq:TransNV}--\eqref{eq:TransAux2}
and dropping terms propotional to $\epsilon^2$ leads to (use $'$ for
$\frac{d}{dx}$)
\begin{eqnarray}
    0&=&4\,\gamma\,f - 4\,f^\pr +f^{\pr\pr\pr}+3\,k^2\,f^\pr\nonumber\\ 
  &&-3\,(u_{0}\,g)^\pr-3\, (v_0\,f)^\pr -3\, u_{0}\,i\,k\,h
   \label{eq:MainNVKexpan}\\
f^\pr &=& +g^\pr-i\,k\,h\label{eq:MainNVKexpanA1}\\
i\,k\,f &=& -h^\pr-i\,k\,g\label{eq:MainNVKexpanA2}
\end{eqnarray}
This is a system of linear ordinary differential equations with
non-constant coefficients.
Use $h(x) =\frac{1}{i\,k}\,(g^\pr(x)-f^\pr(x))$ and the above can
be written in the form
\begin{eqnarray}
  f^{\pr\pr\pr}(x)&=&(-4\,\gamma+3\,u_0^\pr(x))\,f(x)+(4-3\,k^2)\,f^\pr(x)+
   \nonumber\\
   && +3\,u_0^\pr(x)\,g(x) + 6\, u_0(x)\,g^\pr(x) 
\label{eq:MainNVKexpanKdV1}\\
  g^{\pr\pr}(x) &=& k^2\,g(x) +f^{\pr\pr}(x)+k^2\,f(x)
\label{eq:MainNVKexpanKdV2}
\end{eqnarray}

\subsection{Known solutions}
Since the $f_c(x-c\,t+s)$ is a solution of the original NV equation
~\eqref{eq:NVreal}--\eqref{Aux2}, we conclude that
$(u_0(x-s),\,v_0(x-s),\,w_0(x-s))$ solve the
system~\eqref{eq:TransNV}--\eqref{eq:TransAux2} and thus their
derivatives must solve the linearized
problem~\eqref{eq:MainNVKexpan}--\eqref{eq:MainNVKexpanA2}.
Since $\frac{d}{dx}\,\sech^2(x) = -2\,\sech^2(x)\,\tanh(x)$ we find
solutions
\begin{equation}
  \label{eq:Known1}
  f(x) = g(x) = \sech^2(x)\,\tanh(x)\;,\qquad h(x)=0\end{equation}
for $k=\gamma=0$ to the
system~\eqref{eq:MainNVKexpanKdV1}--\eqref{eq:MainNVKexpanKdV2}.
For $k=1$ and $\gamma=0$ we found
\begin{eqnarray}
  \nonumber
  f(x) &=& +\sech^3(x)\\
 g(x) &=& -\sech(x)\,\tanh^2(x)\label{eq:Known2}\\
h(x)&=&-i\,\sech(x)\,\tanh(x)\nonumber
\end{eqnarray}
as a second, explicit solution
of~\eqref{eq:MainNVKexpan}--\eqref{eq:MainNVKexpanA2}.

With the help of these solutions we will construct unstable solution
to the Novikov-Veselov equations, as function of the parameters $k$
and~$\gamma$. Thus we need nonzero solutions of~\eqref{eq:MainNVKexpanKdV1}--\eqref{eq:MainNVKexpanKdV2}.

\subsection{Locating unstable solutions}
Using the function $u_0(x) = -2\,\sech^2(x)$, a matrix notation
translates the high order
system~\eqref{eq:MainNVKexpanKdV1}--\eqref{eq:MainNVKexpanKdV2} into a
system of first order ordinary differential equations.
\[ \frac{d}{dx}\;\Vec{f\\f^\pr\\f^{\pr\pr}\\g\\g^\pr}\,=\,
\Matrix{5}{0&1&0&0&0\\0&0&1&0&0\\
  -4\,\gamma+3\,u_0^\pr&(4-3\,k^2)&0&3\,u_0^\pr(x)&6\,u_0\\
  0&0&0&0&1\\k^2&0&1&k^2&0}\;
\Vec{f\\f^\pr\\f^{\pr\pr}\\g\\g^\pr}\]
\subsubsection{Behavior of solution for large $|x|$}
For large values of $|x|$ we use $u_0(x)\approx 0$ and arrive at a
decoupled system of linear differential equations with constant coefficients.
\begin{eqnarray}
   \frac{d}{dx}\;\Vec{f\\f^\pr\\f^{\pr\pr}}
&=&\Matrix{3}{0&1&0\\0&0&1\\
  -4\,\gamma&(4-3\,k^2)&0\\}\;\Vec{f\\f^\pr\\f^{\pr\pr}}
  \label{eq:FarLimit1}\\
   \frac{d}{dx}\;\Vec{g\\g^\pr}
&=&\Matrix{2}{0&1\\k^2&0}\;\Vec{g\\g^\pr}+\Vec{0\\k^2\,f^\pr+f^{\pr\pr}}
  \label{eq:FarLimit2}
\end{eqnarray}
The characteristic equation of the matrix in~\eqref{eq:FarLimit1} is
given by
\begin{equation}
  \label{eq:characteristic}
  \lambda^3 + (3\,k^2-4)\,\lambda+4\,\gamma=0,
\end{equation}
and we will denote the eigenvalues $\lambda_1, \lambda_2$, and $\lambda_3$ of this system by $p_1, p_2$, and $p_3$, respectively.  
The inhomogeneous system in~\eqref{eq:FarLimit2} leads to eigenvalues
$\lambda_{4,5}=\pm k$.
Using Cardano's formulas for zeros of polynomials of degree 3, we verify that~\eqref{eq:characteristic} has three
distinct real values if
\[4\,\gamma^2 +\frac{(3\,k^2-4)^3}{27}<0.\]
For our domain of $(k,\gamma)$ to be examined we may use
\[ p_3\leq 0\leq p_1,\, p_2\]
and for small values of $|\gamma|$ we find
\begin{equation}
  \label{eq:lambdas}
    \lambda(\gamma)=  \Cases{
   p_1\,\approx\,0+\frac{4}{4-3\,k^2}\,\gamma&\\
   p_2\,\approx\,+\sqrt{4-3\,k^2}-\frac{2}{4-3\,k^2}\,\gamma&\\
   p_3\,\approx\,-\sqrt{4-3\,k^2}-\frac{2}{4-3\,k^2}\,\gamma &}
\end{equation}
with the solution of the form $f(x)\approx c_1e^{p_1 x} + c_2 e^{p_2 x} + c_3 e^{p_3 x}$.
We examine $\gamma\geq 0$, and for $x\to +\infty$ we require the
solution $f(x)$ of~\eqref{eq:FarLimit1} to be bounded, and thus
\[f(x) \approx f_+(x)= c_3\,e^{p_3\,x}.\]
Next, examine~\eqref{eq:FarLimit2} in the form
\[ g_+^{\pr\pr}(x) -k^2\,g_+(x) = k^2\,f_+(x)+f_+^{\pr\pr}(x)
 = c_3\,(k^2+p_3^2)\,e^{p_3\,x}\]
with the solution
\[   g_+(x) = \beta_1\,e^{+k\,x}+ \beta_2\,e^{-k\,x}
   + c_3\,\frac{k^2+p_3^2}{p_3^2 - k^2}\,e^{p_3\,x}.\]
Since this solution has to remain bounded as $x\to+\infty$, we find
\[   g_+(x) = \beta_2\,e^{-k\,x}+ c_3\,\frac{k^2+p_3^2}{p_3^2 -
  k^2}\,e^{p_3\,x}.\]
The case $\gamma\geq 0$ and $x\to-\infty$,
leads in an analogous way to
\[f(x) \approx f_-(x)= c_1\,e^{p_1\,x}+c_2\,e^{p_2\,x}\]
and
\[  g(x)\approx  g_-(x) = \beta_1\,e^{+k\,x}
  + c_1\,\frac{k^2+p_1^2}{p_1^2 - k^2}\,e^{p_1\,x}
  + c_2\,\frac{k^2+p_2^2}{p_2^2 - k^2}\,e^{p_2\,x}.\]
For $0<k^2\ll1$ we find
  \begin{equation}
    \label{eq:trouble}
     p_1^2 \approx\frac{4^2}{(4-3\,k^2)^2}\,\gamma^2 \approx\gamma^2
  \end{equation}
and thus the above formula for $g_(x)$ may not be valid for
$\gamma^2\approx k^2\ll1$~.
For $k=0$ the equation for $g$ simplifies to $g^{\pr\pr}(x) =
  f^{\pr\pr}(x)$, and the conditions at $x=\pm\infty$ imply $g(x)=f(x)$.

\subsubsection{Solutions for intermediate $|x|$.}
Choosing a large value for $M>0$, we construct nontrivial
solutions to~\eqref{eq:MainNVKexpanKdV1}--\eqref{eq:MainNVKexpanKdV2}  for $-\infty<x\leq -M$, then for $-M\leq x\leq+M$, and then
for $+M\leq x<+\infty$. We seek nonzero values of the parameters
$(c_1,\,c_2,\,c_3,\,\beta_1,\,\beta_2)$, such that we find a nonzero
solution.

Use the solutions $f_-$ and $g_-$ to define a matrix
\[\mathbf{T}_-=
\Matrix{5}{
e^{-p_1\,M}&e^{-p_2\,M}&0&0&0\\
p_1\,e^{-p_1\,M}&p_2\,e^{-p_2\,M}&0&0&0\\
p_1^2\,e^{-p_1\,M}&p_2^2\,e^{-p_2\,M}&0&0&0\\
\frac{p_1^2+k^2}{p_1^2-k^2}\,e^{-p_1\,M}&
\frac{p_2^2+k^2}{p_2^2-k^2}\,e^{-p_2\,M}&0&e^{-k\,M}&0\\
\frac{(p_1^2+k^2)\,p_1}{p_1^2-k^2}\,e^{-p_1\,M}&
\frac{(p_2^2+k^2)\,p_2}{p_2^2-k^2}\,e^{-p_2\,M}&0&k\,e^{-k\,M}&0}\]
 with
\[\mathbf{T}_- \Vec{c_1\\c_2\\c_3\\\beta_1\\\beta_2} = 
\Vec{f(-M)\\ f^\pr(-M)\\ f^{\pr\pr}(-M)\\g(-M)\\ g^\pr(-M)}\]
Use the solutions $f_+$ and $g_+$ to define a matrix
\[\mathbf{T}_+=
\Matrix{5}{
0&0&e^{p_3\,M}&0&0\\
0&0&p_3\,e^{p_3\,M}&0&0\\
0&0&p_3^2\,e^{p_3\,M}&0&0\\
0&0&\frac{p_3^2+k^2}{p_3^2-k^2}\,e^{p_3\,M}&0&e^{-k\,M}\\
0&0&\frac{(p_3^2+k^2)\,p_3}{p_3^2-k^2}\,e^{p_3\,M}&0&-k\,e^{-k\,M}}\]
with
\[\mathbf{T}_+ \Vec{c_1\\c_2\\c_3\\\beta_1\\\beta_2} = 
\Vec{f(+M)\\ f^\pr(+M)\\ f^{\pr\pr}(+M)\\g(+M)\\ g^\pr(+M)}\]
Then use an ODE solver to examine the
system~\eqref{eq:MainNVKexpanKdV1}--\eqref{eq:MainNVKexpanKdV2} on the interval
$[-M,M]$ as an initial value problem. This leads to a matrix $\mathbf{T}$ such
that
\[\mathbf{T} \Vec{f(-M)\\ f^\pr(-M)\\ f^{\pr\pr}(-M)\\g(-M)\\ g^\pr(-M)}
 = \Vec{f(+M)\\ f^\pr(+M)\\ f^{\pr\pr}(+M)\\g(+M)\\ g^\pr(+M)}\]
Now we have two methods to compute the values of the solution at
$x=+M$, and the system~\eqref{eq:MainNVKexpanKdV1}--\eqref{eq:MainNVKexpanKdV2} 
has a nonzero solution if and only if
\begin{equation}
  \label{eq:NVODEcond}
  D(k,\gamma)\,=\,\det(\mathbf{M}(k,\gamma)) =
\det(\mathbf{T}\cdot \mathbf{T}_--\mathbf{T}_+)=0
\end{equation}
Thus we examine solutions of this equation as functions of the
parameters $k$ and~$\gamma$.

\subsection{The special case $k=0$}
For $k=0$, equation \eqref{eq:MainNVKexpanKdV2} reads as
$g^{\pr\pr}(x)=f^{\pr\pr}(x)$ and the only solution satisfying
$g(\pm\infty)=f(\pm\infty)$ is $g(x)=f(x)$. Then~\eqref{eq:MainNVKexpanKdV1}
leads to
\begin{equation}
  \label{eq:MainExpandKdV1k0}
\frac{d}{dx}\,\Vec{f\\f^\pr\\f^{\pr\pr}} =
\Matrix{3}{0&1&0\\0&0&1\\
-4\,\gamma+6\,u_0^\pr&4+6\,u_0&0}
\;\Vec{f\\f^\pr\\f^{\pr\pr}}
\end{equation}
Using the same approach as in the previous section we define
\begin{eqnarray*}
  \mathbf{T}_-\,\Vec{c_1\\c_2\\c_3} =  \Vec{f(-M)\\f^\pr(-M)\\f^{\pr\pr}(-M)}
&=&\Matrix{3}{e^{-p_1\,M}&e^{-p_2\,M}&0\\p_1\,e^{-p_1\,M}&p_2\,e^{-p_2\,M}&0\\
p_1^2\,e^{-p_1\,M}&p_2^2\,e^{-p_2\,M}&0}\,\Vec{c_1\\c_2\\c_3}\\
  \mathbf{T}_+\,\Vec{c_1\\c_2\\c_3} =  \Vec{f(+M)\\f^\pr(+M)\\f^{\pr\pr}(+M)}
&=&\Matrix{3}{0&0&e^{p_3\,M}\\0&0&p_3\,e^{p_3\,M}\\0&0&p_3^2\,e^{p_3\,M}}
  \,\Vec{c_1\\c_2\\c_3}\\
\mathbf{T}_0 \Vec{f(-M)\\ f^\pr(-M)\\ f^{\pr\pr}(-M)}
 &=& \Vec{f(+M)\\ f^\pr(+M)\\ f^{\pr\pr}(+M)}
\end{eqnarray*}
where $\mathbf{T}_0$ is constructed using an ODE solver
for~\eqref{eq:MainExpandKdV1k0}.
Then the system~\eqref{eq:MainNVKexpanKdV1}--\eqref{eq:MainNVKexpanKdV2} 
has a nonzero solution for $k=0$ if and only if
\begin{equation}
  \label{eq:NVODEcond0}
  D_0(\gamma)\,=\,\det(\mathbf{M}_0(\gamma)) =
\det(\mathbf{T}_0\cdot \mathbf{T}_--\mathbf{T}_+)=0
\end{equation}

\section{Numerical Results on the Instabilities of Plane-Wave 
Soliton Solutions}

\subsection{Locating unstable solutions}
\begin{figure}[htbp]
  \begin{center}
    \leavevmode
   \includegraphics[width=12cm]{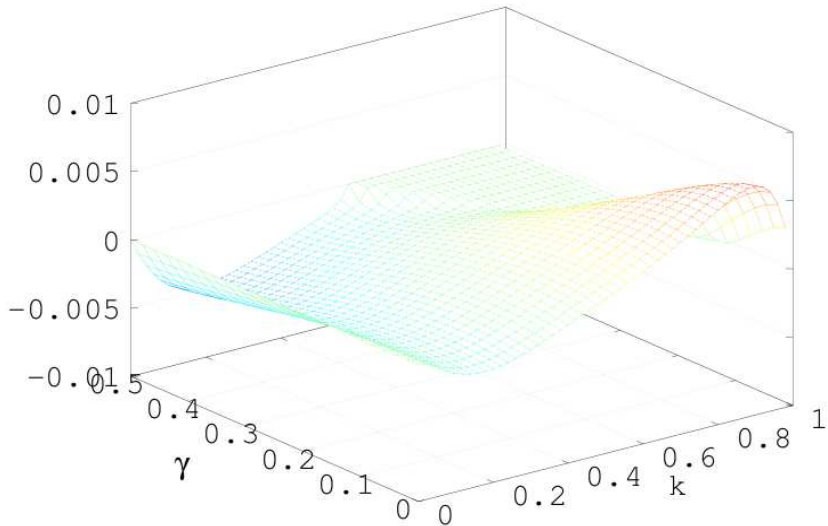}
    \caption{The real part of $\det(\mathbf{M})$ as a function of $k$
      and $\gamma$}
    \label{fig:DetT}
  \end{center}
\end{figure}

Based on expression~\eqref{eq:NVODEcond} we generate plots of the
function $D(k,\gamma)= \det(\mathbf{M}(k,\gamma))$ on a domain $0\leq
k\leq 1$ and $0\leq\gamma\leq 0.5$, leading to
Figure~\ref{fig:DetT}. In the corner $k\approx 1$ and
$\gamma\approx 0.5$ the real part of the function vanishes, but a
second plot verifies that the imaginary part is different from
zero. Thus Figure~\ref{fig:DetT} indicates that we have a clearly
defined solution curve of $\det(\mathbf{M}(k,\gamma))=0$, away from
the origin.

\begin{figure}[htbp]
  \begin{center}
    \leavevmode
\subfigure[$\det\,\mathbf{M}(k,\gamma)$ close to origin]{\label{fig:DetTSmall}
   \includegraphics[width=7cm]{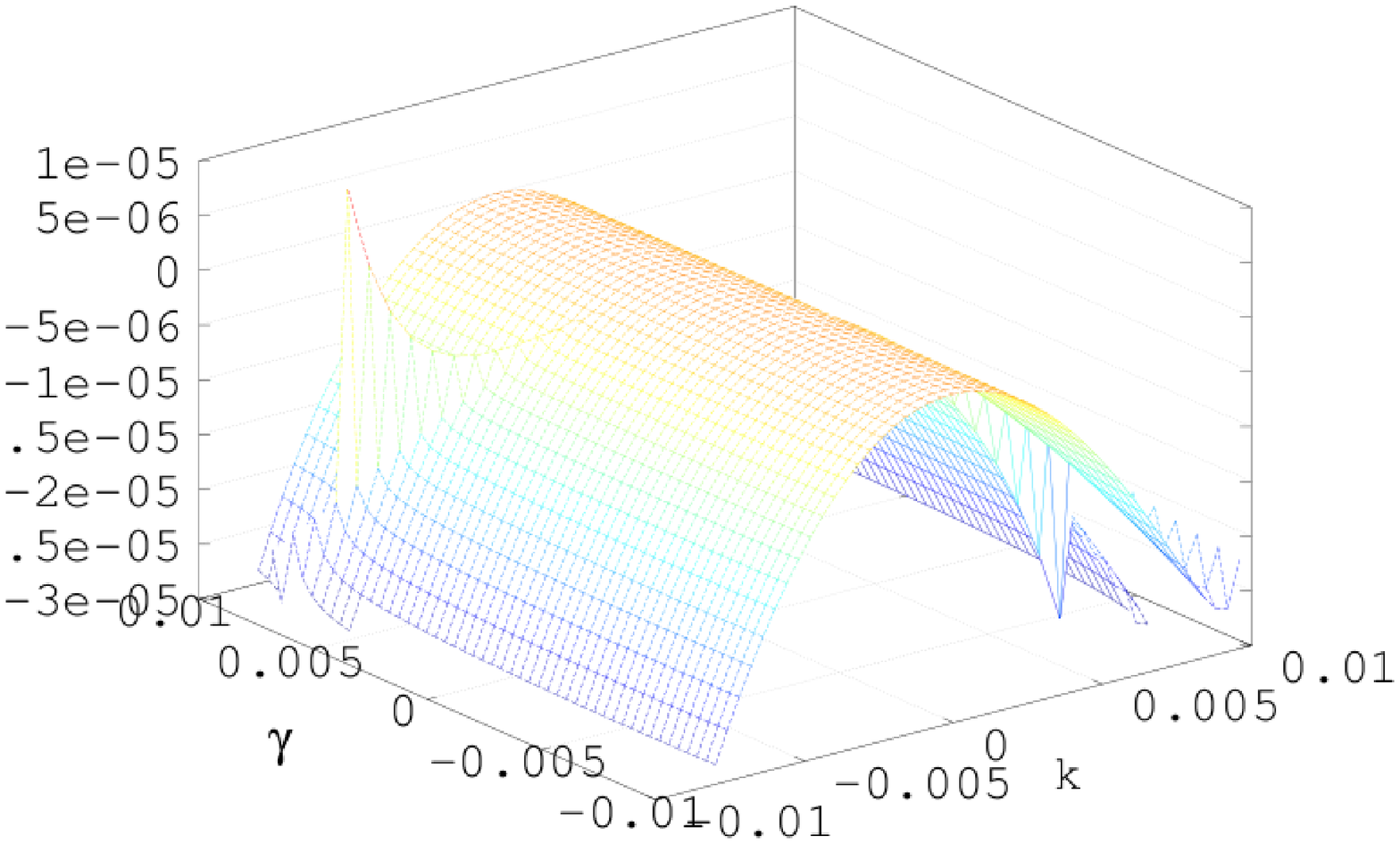}}\hfil
\subfigure[$\det\,\mathbf{M}_0(\gamma)$]{\label{fig:DetTk0asFuncOfga}
   \includegraphics[width=6cm]{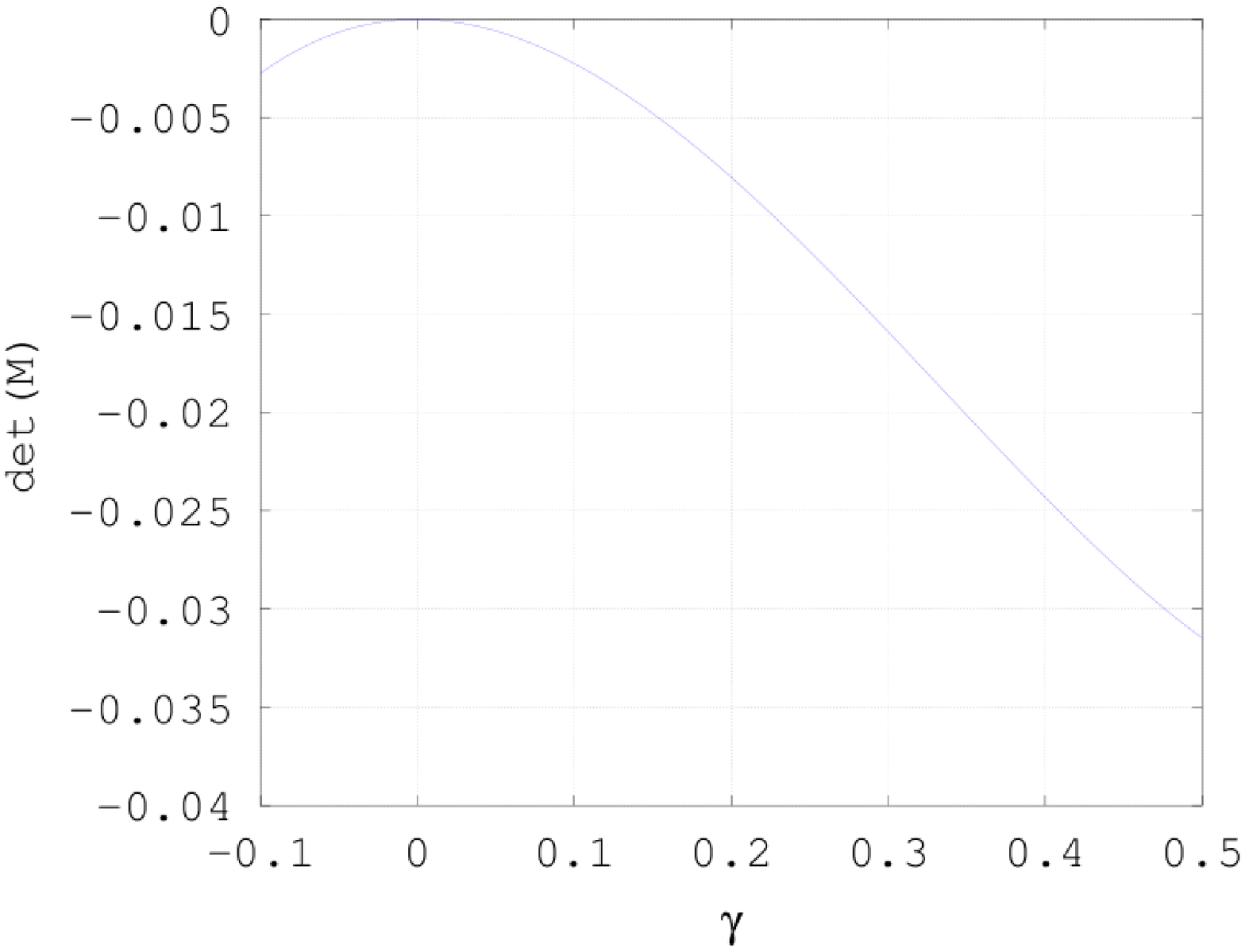}}
    \caption{Behavior at the origin and along the axis $k=0$}
    \label{fig:criticaValues}
  \end{center}
\end{figure}

Since the behavior close to $(k,\gamma)\approx(0,\,0)$ is critical, we
examine this section with a finer resolution, leading to
Figure~\ref{fig:DetTSmall}. The obvious spikes are caused by the
zeros in the denominator in condition~\eqref{eq:trouble}.
Figure~\ref{fig:DetTSmall} suggests the existence of a solution
along the axis $k=0$. Using~\eqref{eq:NVODEcond0} we generate
Figure~\ref{fig:DetTk0asFuncOfga}. As a consequence
the only solution along the axis $k=0$ is at $\gamma=0$~. 
Thus the known solution~\eqref{eq:Known1} is an isolated solution in
the parameter space $(k,\gamma)$ at $(0,\,0)$.

With the above preparation we can now construct values of $(k,\gamma)$
leading to nonzero solutions
of~\eqref{eq:MainNVKexpanKdV1}--\eqref{eq:MainNVKexpanKdV2} and thus
for $\gamma>0$ to unstable soliton solutions of the NV
equations~\eqref{eq:NVreal}--\eqref{Aux2}. 

\begin{figure}[htbp]
  \begin{center}
    \leavevmode
\subfigure[solution curve of $|\det(\mathbf{M})|=0$]{\label{fig:SolveOnCurve12A}
      \includegraphics[width=6cm]{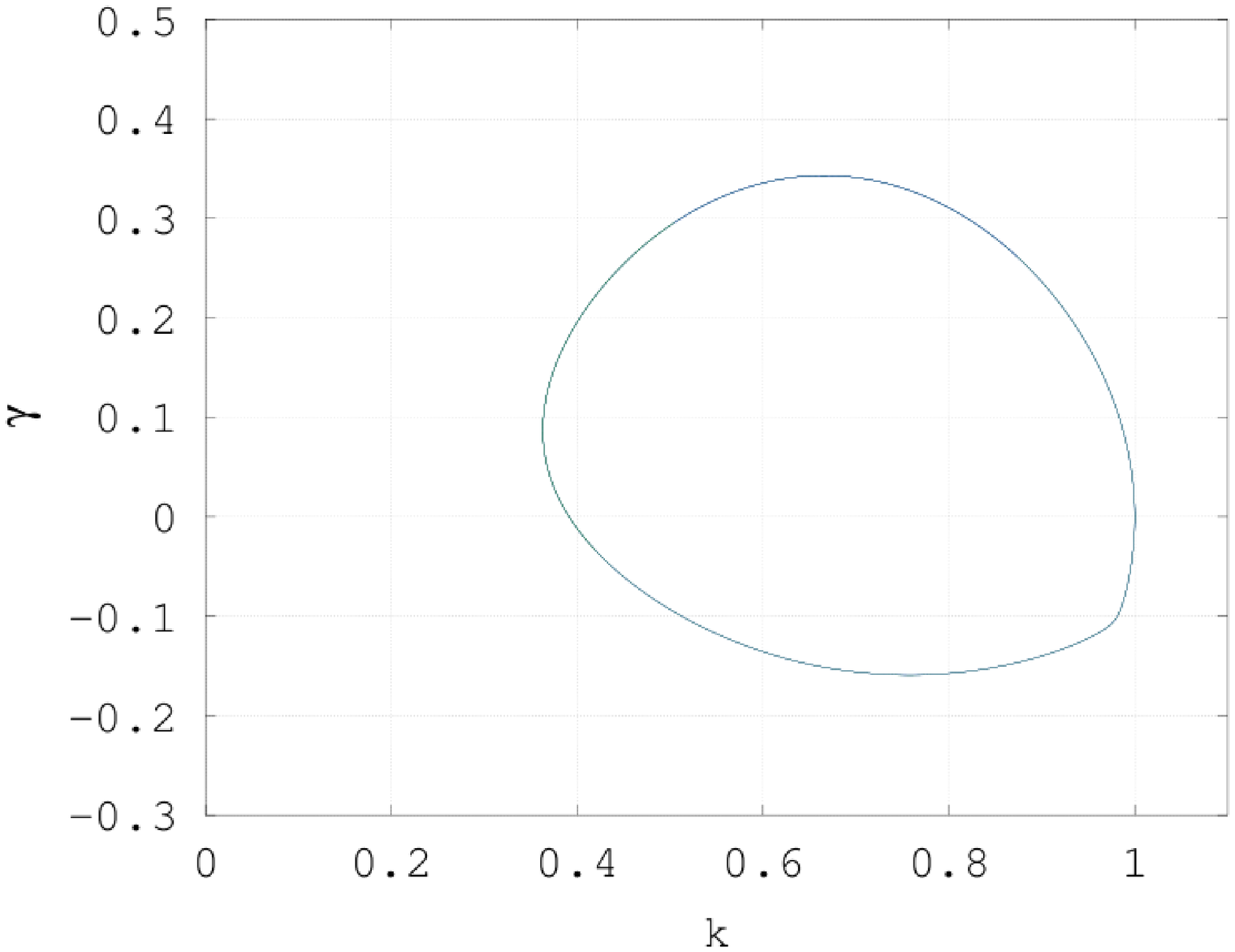}} \hfil
\subfigure[parameters leading to unstable solutions]{\label{fig:SolveOnCurve12B}
      \includegraphics[width=6cm]{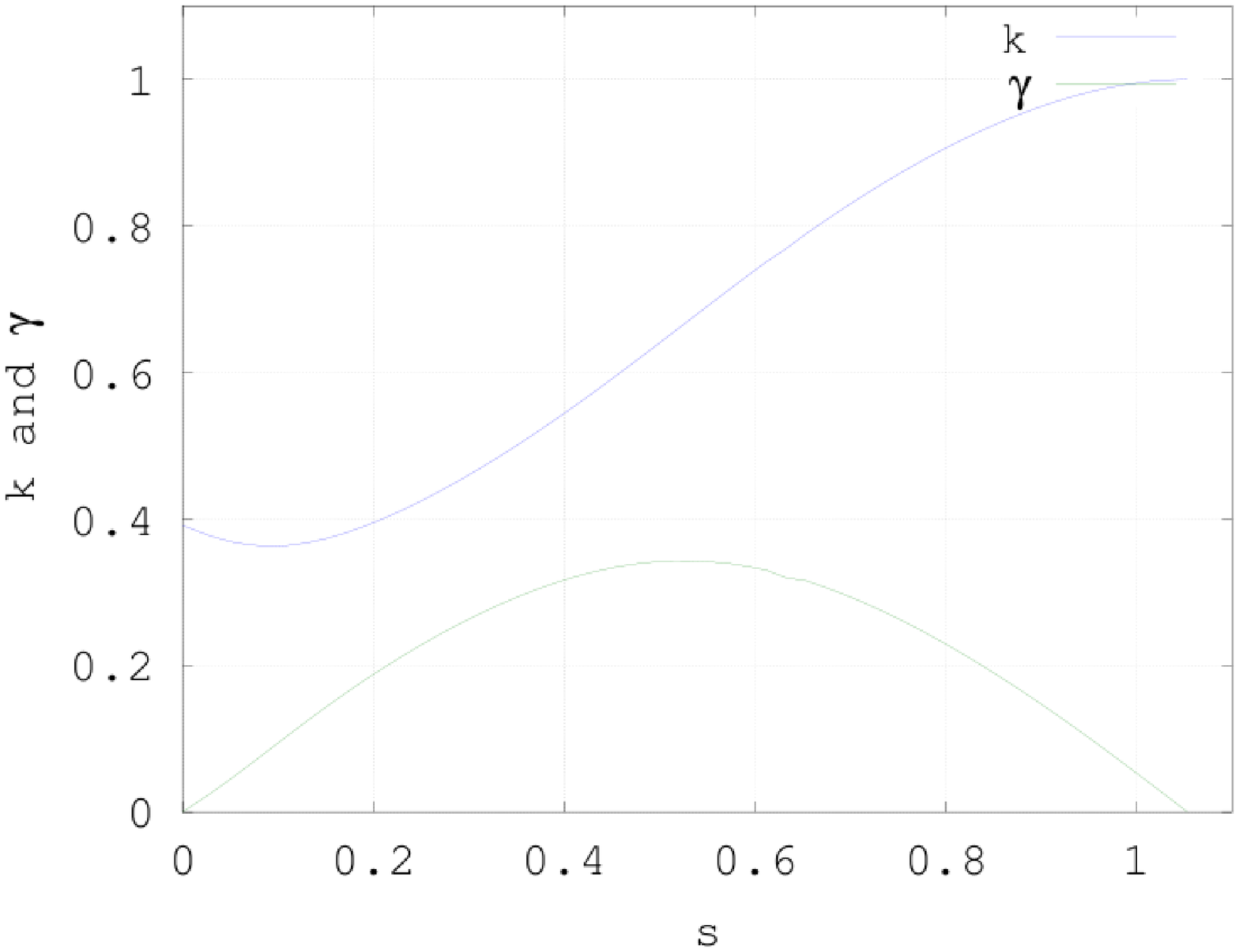}}
    \caption{Solution curve of $|\det(\mathbf{M})|=0$ and parameters
      $k$ and $\gamma$ leading to unstable solutions} 
    \label{fig:SolveOnCurve12}
  \end{center}
\end{figure}

We trace a solution curve of~\eqref{eq:NVODEcond} by generating an arc
length parametrization of the curve. Using an arbitrary initial point
on the curve (use a contour plot of Figure~\ref{fig:DetT}) and a
starting direction, we minimize $|\det(\mathbf{M}(k,\gamma))|$ along a
straight line segment orthogonal to the stepping direction. Using this
local minimum, we adjust the stepping direction and then make a small
step to follow the solution curve. While stepping along the curve we
verify that we actually have a solution of~\eqref{eq:NVODEcond}, and
not only a minimum. This proved to be a stable algorithm and generated
Figure~\ref{fig:SolveOnCurve12}.

Observe that Figure~\ref{fig:SolveOnCurve12A} also displays negative
values of~$\gamma$, which do not lead to unstable solutions
of~\eqref{eq:NVreal}--\eqref{Aux2}. We display these values to
confirm that we have a closed curve without branching points.

For the parameter values of $(k,\gamma)$ in Figure~\ref{fig:SolveOnCurve12B}
there are nonzero functions $f(x)$, $g(x)$ and $h(x)$, such that for
small $\epsilon$ we have solutions of the
equations~\eqref{eq:TransNV}--\eqref{eq:TransAux2} of the form 
\eqref{eq:Main1}--\eqref{eq:Main3}. Since the $y$ dependence of these
functions is of the form $e^{i\,k\,y}$, the functions are periodic in $y$
with a period of $L_y = \frac{2\,\pi}{k}$. The corresponding exponent
is shown in Figure~\eqref{fig:SolveOnCurve12B}.
Thus the Novikov--Veselov equations~\eqref{eq:NVreal}--\eqref{Aux2}
with initial condition $u(x,y,0) = -2\,c\,\sech^2(\sqrt{c}\,x)$ will
lead to an unstable soliton solution.

\begin{theorem} \label{thm:3}
The soliton solutions
$u(x,y,t)=v(x,y,t)=-2\,c\,\sech^2(\sqrt{c}\,(x-c\,t))$ and $w(x,y,t)=0$
of the NV equations~\eqref{eq:NVreal}--\eqref{Aux2} are not stable.
For values of $0.363<k\leq 1$ there are $y$--periodic, unstable
contributions with a period of $L_y = \frac{2\,\pi}{k\,\sqrt{c}}$.
\end{theorem}

\subsection{Constructing unstable solutions numerically}
\begin{figure}[htbp]
  \begin{center}
    \leavevmode
\subfigure[shape of the solution]{\label{fig:Sol1A}
      \includegraphics[width=7cm]{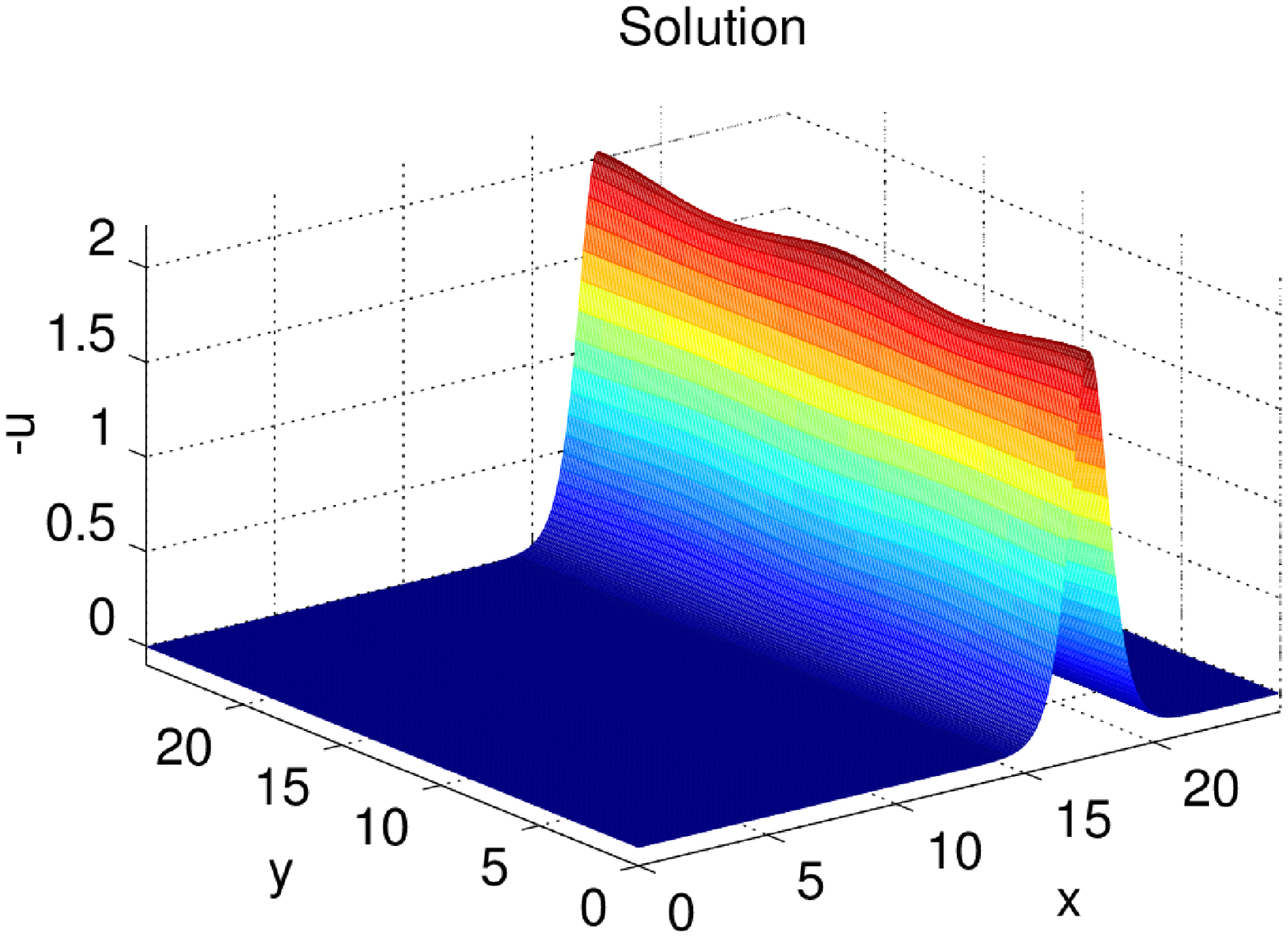}}
\subfigure[shape of the perturbation]{\label{fig:Sol1B} \hfil
      \includegraphics[width=7cm]{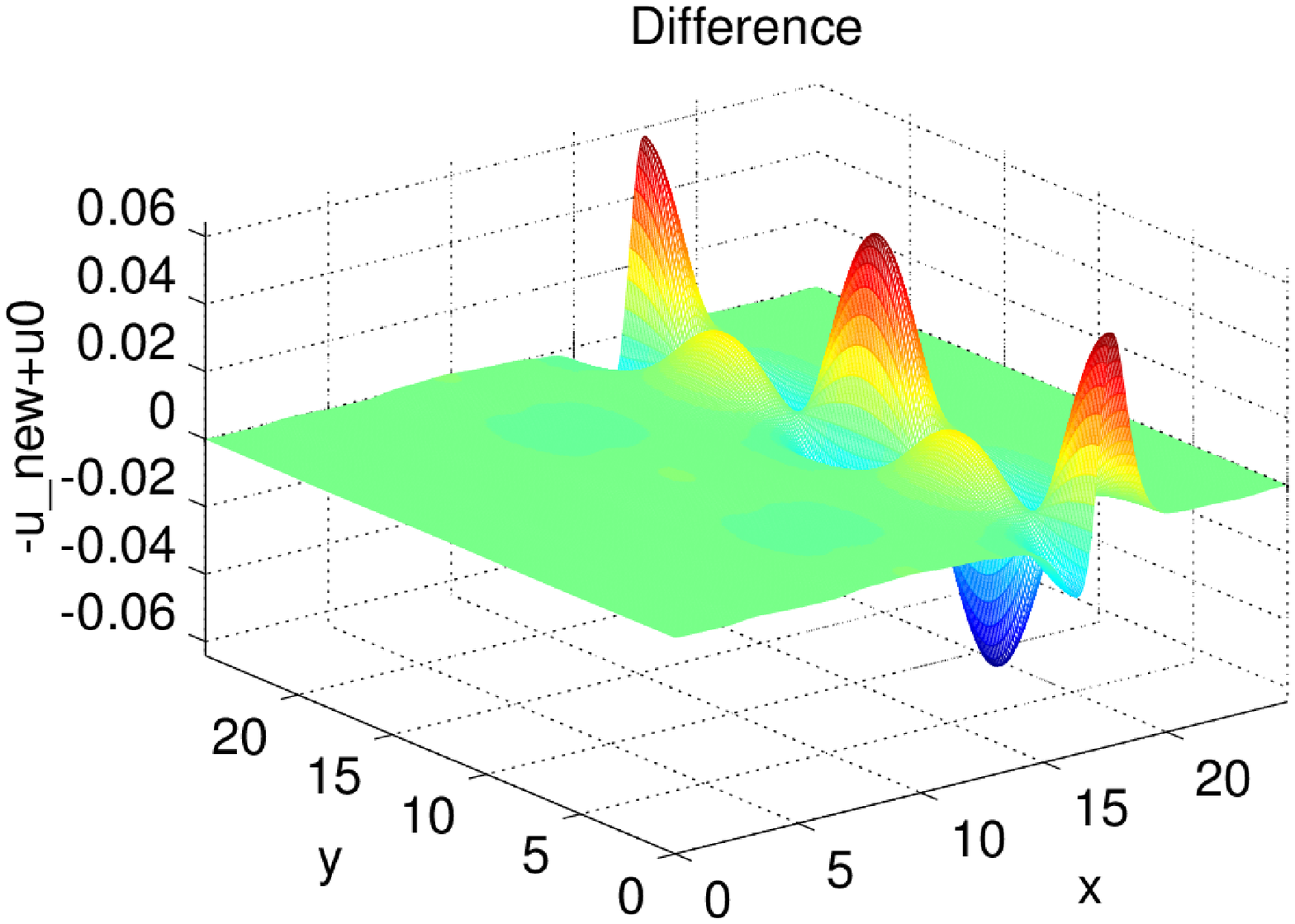}}
    \caption{A perturbed solution at time 5}
    \label{fig:Sol1}
  \end{center}
\end{figure}

\begin{figure}[htbp]
  \begin{center}
    \leavevmode
\subfigure[at $k\approx0.5$ and
$\gamma\approx0.3$]{\label{fig:Sol1ShapePerturbA}
      \includegraphics[width=6cm]{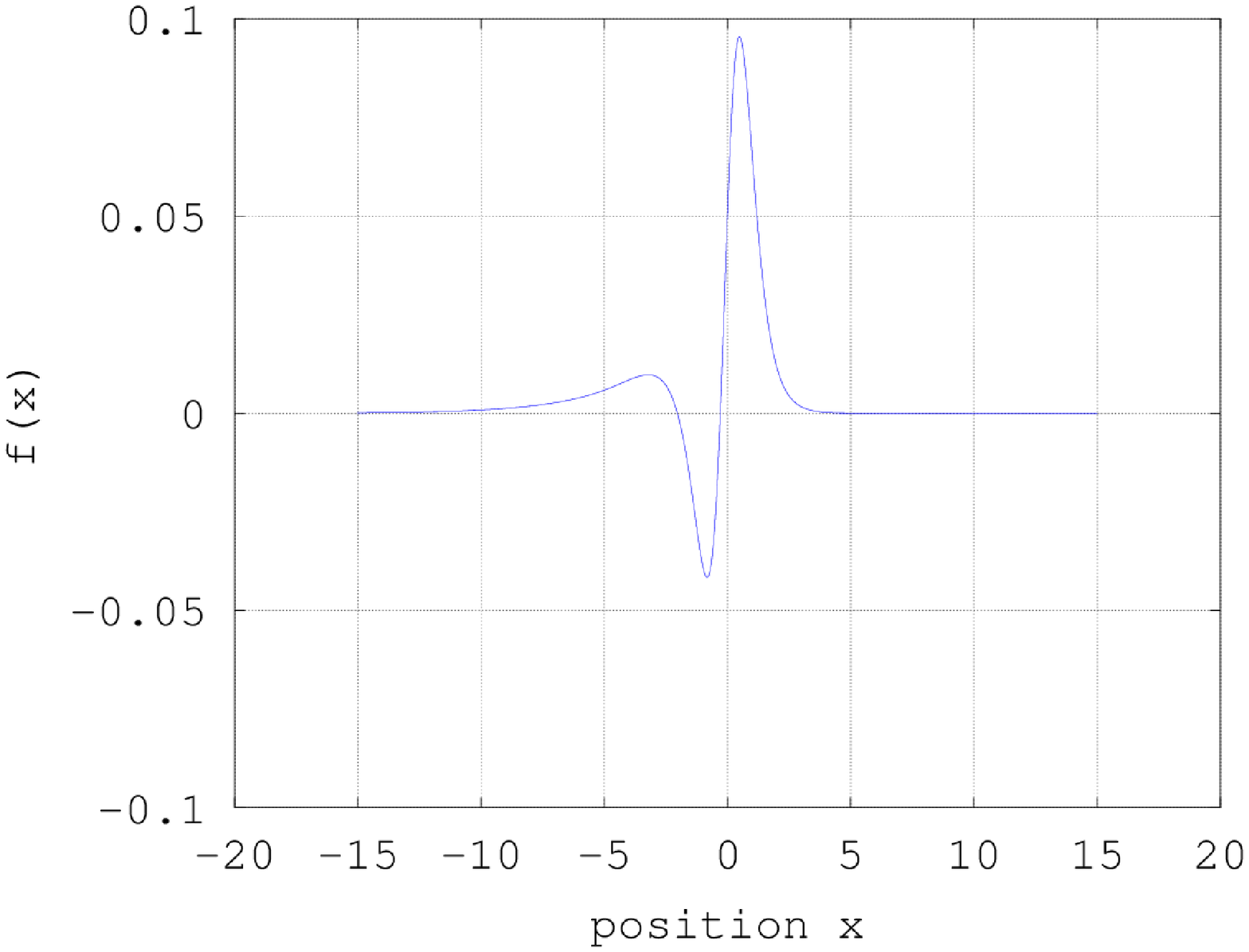}}
\subfigure[as function of arc length $s$]{\label{fig:Sol1ShapePerturbB} \hfil
      \includegraphics[width=6cm]{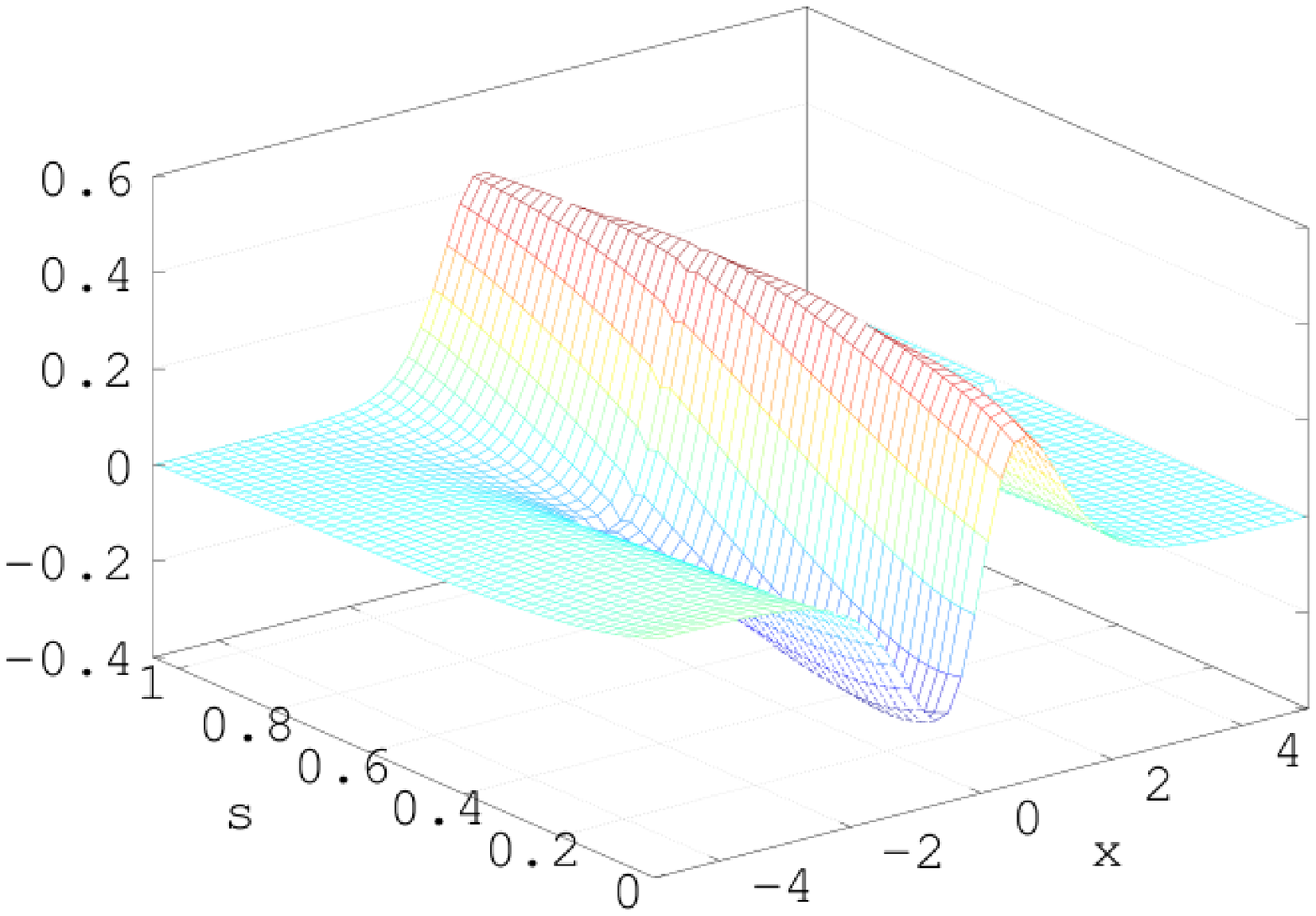}}
    \caption{Shape of the perturbations $f(x)$}
    \label{fig:Sol1ShapePerturb}
  \end{center}
\end{figure}

As an example, in this section we construct one of the above unstable
solutions numerically, using the algorithm from
Section~\ref{sec:Spectral} for periodic solutions in $x$ and
$y$. Since the soliton solution $\sech^2(x)$ decays rapidly, the above
results still apply when working on a sufficiently large domain.
The steps of the algorithm are as follows:
\begin{enumerate}
\item Choose values of $(k,\gamma)$ in
  Figure~\ref{fig:SolveOnCurve12B} and determine the eigenvector
  $(c_1,c_2,c_3,\beta_1,\beta_2)$ for the zero eigenvalue.
\item Use the algorithm leading to the matrices $\mathbf{T}$,
  $\mathbf{T}_-$ and $\mathbf{T}_+$ to construct the nonzero functions
  $f(x),g(x),h(x)$.
\item Pick a size domain such that $\frac{2\,\pi}{k}$ periodic
  functions in $y$ are admissable.
\item Construct initial values, using~\eqref{eq:Main1R}--\eqref{eq:Main3R},
  use a small value of $\epsilon$. 
  Without taking the transformations~\eqref{eq:Transformations} into account
  we obtain a speed of $c=1$ of the unperturbed soliton.
\item Solve the NV equations~~\eqref{eq:NVreal}--\eqref{Aux2}, using
  the algorithm presented in Section~\ref{sec:Spectral}.
\item The deviation from the single soliton solution should not change
  its shape, but the size is expected to be proportional to $e^{\gamma\,t}$.
\item Solitions for speeds $c\neq 1$ can be constructed similarly, using the
  transformations~\eqref{eq:Transformations}.
\end{enumerate}
The evolution of a perturbed soliton from Theorem \ref{thm:3} was
computed using the semi-implicit pseudo-spectal method.  Here we chose
$k=0.504$ and $\gamma=0.296$, and the graph of the perturbation $f(x)$
is found in Figure~\ref{fig:Sol1ShapePerturbA}. 
As initial value we chose a perturbed KdV solition with speed $c=1$, starting
at $x=10$. Find the solution and the difference to the unperturbed KdV soliton
at time~$5$ in Figure~\ref{fig:Sol1}. The corresponding animations
are available on the web site~\cite{webNV}. The exponential growth of the
perturbation with exponent $\gamma$ is numerically confirmed.

One can construct the shapes of the functions $f(x)$ for all positive values
of $\gamma$ along the arc in Figure~\ref{fig:SolveOnCurve12} to obtain
Figure~\ref{fig:Sol1ShapePerturbB}. The solutions constructed for
$(k,\gamma)=(1,\,0)$ have to match the known exact
solutions~\eqref{eq:Known2}, which is confirmed.

\section{Conclusions}

In this work a semi-implicit pseudo-spectral method was introduced for
the numerical computation of evolutions of solutions to the NV
equation, constituting the first numerical implementation of a
spectral method for a system of soliton nonlinear PDE's.  A linear
stability analysis yields a stability condition for the Crank-Nicolson
scheme on the linearized IBVP.  The instability of traveling wave
solutions to transverse perturbations was established by the 
{\bf K}-expansion method, and unstable soliton solutions were
constructed. The evolution of an example was computed numerically by
the semi-implicit pseudo-spectral method.


\bibliographystyle{amsplain}
\bibliography{NVStability_arxiv}

\end{document}